\documentstyle[12pt,preprint,aps,epsfig]{revtex}
\draft
\title{Collective Deceleration of Ultrarelativistic Nuclei\\
and Creation of Quark-Gluon Plasma}
\author{I. N. Mishustin$^{a,b,c}$ and J. I. Kapusta$^a$}
\address{
 $^a$School of Physics and Astronomy, University of Minnesota,
Minneapolis, MN 55455, USA\\
 $^b$Niels Bohr Institute, Blegdamsvej 17, DK-2100 Copenhagen {\O}, Denmark\\
 $^c$Kurchatov Institute, Russian Research Center, 123182 Moscow, Russia}

\begin{document}

\preprint{ NUC-MINN-14/01-T}

\maketitle
\begin{abstract}
We propose a unified space-time picture of baryon stopping and quark-gluon 
plasma creation in ultrarelativistic heavy-ion collisions. It is assumed that 
the highly Lorentz contracted nuclei are decelerated by the coherent color field
which is formed between them after they pass through each other. This process 
continues until the field is neutralized by the Schwinger mechanism. 
Conservation of energy and momentum allow us to calculate the energy losses of 
the nuclear slabs and the initial energy density of the quark-gluon plasma.      
\end{abstract}

Ultrarelativistic heavy-ion collisions offer a unique possibility to study 
collective dynamical effects at the partonic level. Before a collision the 
partons are confined in coherent field configurations and the Lorentz contracted 
nuclei can propagate in the physical vacuum without distortion. After the 
nuclei collide, thousands of partons are quickly liberated. The nuclei are 
transformed into two partonic sheets which recede from each other, leaving 
behind the perturbative QCD vacuum and strong gluon fields \cite{mclerran}. 
Since color charges on the sheets are distributed stochastically, they generate 
chromoelectric fields which are nonuniform in the transverse plane. This field 
configuration may be envisaged as a collection of 
densely packed color flux tubes or strings stretched between the sheets. The 
energy accumulated in the coherent field is taken from the kinetic energy of the 
receding partonic sheets that causes their deceleration. At later times 
the coherent fields are neutralized via the Schwinger pair-production 
mechanism \cite{schwinger}. 

In this letter we formulate a simple model to describe the collective deceleration 
of ultrarelativistic nuclei by the coherent color field. This model is similar in 
spirit to that proposed recently in ref. \cite{csernai}. Some geometrical and kinematical 
aspects of the model are also similar to those introduced in ref. \cite{ivanov}. 
All calculations below are performed in the center-of-rapidity frame where the colliding 
nuclei have initial rapidities of $\pm y_0$.

We consider only beam energies so great that the nuclei can be thought of as 
very thin, Lorentz contracted, sheets.  Each sheet is divided into many small 
elements, ``slabs'', of unit transverse area labeled by an index $a$ where $a=p$ 
for the projectile nucleus and $a=t$ for the target nucleus. Each slab is 
characterized by a baryon number $N_{a}$ which is assumed to be strictly 
conserved. This number can be expressed through the slab thickness, $l_a$,
in the rest frame of the respective nucleus,
\begin{equation}
N_{a}({\bf b})=\rho_0l_{a}({\bf b})= \int dz \rho_{a}({\bf b},z)~,
\end{equation}  
where $\rho_a({\bf b},z)$ is the density distribution in nucleus $a$, 
${\bf b}$ is the position of the slab in the impact parameter plane, and
$\rho_0\approx 0.15$ fm$^{-3}$ is the equilibrium nuclear density. 
For a uniform density distribution $l(b)=2\sqrt{R^2-b^2}$, where $R=r_0A^{1/3}$ is the
nuclear radius and $r_0=\left(4\pi\rho_0/3\right)^{-1/3}\approx$ 1.17 fm.
We assume that before and after nuclear overlap, which takes place at $t=0$, each slab 
propagates as a rigid body along the beam axes $z$. The energy 
and momentum of slab $a$ is parameterized in terms of its proper energy per baryon 
$\varepsilon_a$ and longitudinal rapidity $y_{a}$ (for details see ref.~\cite{ivanov}), 
\begin{equation}
E_{a}= N_{a}\varepsilon_{a}\cosh{y_{a}}~,~~~
P_{a}= N_{a}\varepsilon_{a}\sinh{y_{a}}~.
\end{equation} 
In general, $\varepsilon_{a}$ may 
differ from the energy per baryon in normal nuclei, $\varepsilon_0\approx m_N$, 
the nucleon mass, due to hard parton production at $t=0$.  

As a result of soft 
parton-parton interactions the slabs acquire color charges which generate the  
chromoelectric field or several strings in the space between them, see Fig.~1(a). 
For simplicity we disregard here a short time delay which is needed for the formation 
of the coherent field. At $t>0$ the trajectories of the projectile and target slabs, 
$z_p(t)$ and $z_t(t)$, are affected by 
the energy and momentum losses for stretching the strings. If the string tension 
(energy per unit length) for individual strings is $\sigma$, and the number of 
strings per unit transverse area is $n$, then the potential energy stored in 
strings is $V(z)=n\sigma |z_p-z_t|$ (for definiteness we assume that $z_p>z_t$).  
Accordingly the force or, in the present 
context, the longitudinal pressure exerted by strings on slab $a$ is 
$-\partial V/\partial z_a=\mp n\sigma$, and therefore Newton's equation of 
motion has the form 
\begin{equation} \label{P}
\frac{dP_{a}}{dt}=\mp n\sigma~.
\end{equation}
From here on the upper and lower signs correspond to $a=p$ and $a=t$, respectively.  
The net force acting on both slabs is of course zero and the total energy of the 
slabs+field system is conserved.
Realizing that the energy lost by a slab while traversing distance $dz_a$ is 
equal to $\mp n\sigma dz_a$, one can also write the energy conservation equation as
\begin{equation} \label{E}
\frac{dE_{a}}{dz_a}=\mp n\sigma~.
\end{equation}

These equations are identical to those describing the motion of massive 
capacitor plates due to the action of a uniform (chromo)electric field of a capacitor. 
This is not surprising because the action of many collinear flux tubes, or 
strings, is equivalent to the action of a uniform field ${\bf E}$
with the same energy density, $\epsilon_f={\bf E}^2/2=n\sigma$. It is important 
to note that $\epsilon_f$ is a Lorentz invariant quantity.

Using eqs. (\ref{P}) and (\ref{E}) and  the definition of slab velocity, 
\begin{equation} \label{v}
v_a=\frac{dz_a}{dt}=\frac{P_a}{E_a}~,
\end{equation}
one finds that $P_adP_a=E_adE_a$. This means that the internal energies of 
the slabs do not change in the course of deceleration, that is
$E_a^2-P_a^2=N_a^2\varepsilon_a^2={\rm constant}$.
The solution to equations (\ref{P}) and (\ref{E}) for the initial condition 
$y_{a}(0)=\pm y_{0}$, $z_{a}(0)=0$ are easily found to be:
\begin{equation} \label{sinh}
\sinh{y_{a}}=\pm \left(\sinh{y_{0}}-\frac{t}{\lambda_a}\right)~,
\end{equation}
\begin{equation} \label{cosh}
\cosh{y_{a}}=\cosh{y_{0}}\mp\frac{z_{a}}{\lambda_a}~,
\end{equation}
where $\lambda_a=\left(\varepsilon_a\rho_0/\epsilon_f\right)l_a$ is the characteristic deceleration length.

Eliminating  $y_{a}$ from these equations one obtains the slab trajectories
\begin{equation}
\left(\cosh{y_{0}}\mp\frac{z_{a}}{\lambda_a}\right)^2-
\left(\sinh{y_{0}}-\frac{t}{\lambda_a}\right)^2=1~.
\end{equation}
These are parts of hyperbolae shown in Fig.~2.  Initially the slab trajectories
are very close to the light cone but later on 
they increasingly deviate from it. If nothing else were to happen, 
the slabs would reach their respective turning points at $t=\lambda_a\sinh{y_{0}}$, 
$z=\pm \lambda_a\left(\cosh{y_{0}}-1\right)$, and then reverse direction. 
As an example, Fig.~2(b) shoes the projectile trajectory with a turning point.  
This is the ``yo-yo'' type of motion well known in string models.  However, such 
dynamics is very unlikely in nuclear collisions because of irreversible 
processes associated with string decay. 

When the strings become long enough quark-antiquark and gluon pairs can be 
produced from vacuum via the Schwinger mechanism \cite{schwinger}. Their color charges will 
screen the chromoelectric fields and eventually neutralize them. For various 
idealized situations these processes were studied intensively during the last 
decade \cite{eskola}. Here we want to avoid numerical 
simulations of the complicated plasma-field dynamics, which is not very well 
understood yet.  Instead we adopt a simplified picture assuming that the strings 
decay suddenly at a certain proper time $\tau=\tau_0\sim1$ fm, where 
$\tau=\sqrt{t^2-z^2}$. As a result, the partonic plasma is created at the hyperbola
$\tau=\tau_0$ (see Fig.~2). As we will see below, even this simplified picture allows us 
to make nontrivial predictions concerning the energy density and velocity field 
of the plasma at $\tau=\tau_0$. 

As shown in Fig.~1(b), the structure of the system changes dramatically at $t>\tau_0$.
Now the chromoelectric fields are absent in the middle but still 
remain in the regions adjacent to the projectile and target slabs. The gap between 
these two regions is filled by the partonic plasma. As time progresses the boundaries 
between the plasma and coherent fields move closer to the slabs, 
and the gap expands.  Since in a time interval $dt$ the boundaries traverse a 
distance $dz$ along the hyperbola $\sqrt{t^2-z^2}=\tau_0$, their velocities are 
$dz/dt=t/\sqrt{t^2-\tau_0^2}>1$. Although this velocity is superluminal,
it has no physical significance since the boundary is not a material object. The 
plasma itself moves in accordance with the laws of special 
relativity. To see this, let us write the energy-momentum tensor of the plasma in the 
standard form
\begin{equation} \label{T}
T^{\mu\nu}=(\epsilon+p)u^\mu u^\nu-pg^{\mu\nu}~,
\end{equation}
where $\epsilon$ and $p$ are the proper energy density and pressure, $u^\mu$ is 
the collective 4-velocity of the plasma, and $g^{\mu\nu}$=diag(1,-1,-1,-1). At 
its creation the plasma is assumed to have negligible  transverse velocity; 
collective transverse expansion develops at later times. Thus the 4-velocity can be chosen 
as $u^{\mu}=\gamma(1,0,0,v)$ where $v$ is the longitudinal velocity and 
$\gamma=1/\sqrt{1-v^2}$. Now we can write equations expressing the conservation 
of energy and momentum across the boundary between the plasma and the 
chromoelectric field. In time $dt$, when the boundary moves a distance $dz$, the 
energy per unit area subtracted from the field is $\epsilon_fdz$. This change must 
be  equal to the energy of a newly produced plasma slice $dz$: 
$T^{00}dz=\epsilon_fdz$. Using eq. (\ref{T}) this gives 
\begin{equation} \label{ed}
(\epsilon+p)\gamma^2-p=\epsilon_f~.
\end{equation}
Since the field exerts a force $n \sigma=\epsilon_f$ on the plasma charges, the change of 
momentum in the plasma slice $dz$ is $T^{03}dz=\epsilon_fdt$. This leads to
\begin{equation} \label{pr}
(\epsilon+p)\gamma^2v=\epsilon_f\frac{dt}{dz}=\epsilon_f\frac{z}{t}~.
\end{equation}   
In the last equality we have used the condition that the boundary moves along 
the hyperbola 
$\sqrt{t^2-z^2}=\tau_0$ so that $tdt=zdz$. The two equations (\ref{ed}) and 
(\ref{pr}) allow us to find two quantities, $\epsilon$ and $v$, characterizing 
the plasma at $\tau=\tau_0$.  Especially simple results are obtained in the case 
of free streaming, $p=0$ (dust equation of state), which seems most appropriate 
for the early stages of the plasma evolution. By dividing eq. (\ref{pr}) by eq. (\ref{ed})
we get
\begin{equation} \label{vp}
v(\sqrt{t^2-z^2}=\tau_0)=\frac{z}{t}=\tan{\eta}~,
\end{equation}
\begin{equation} \label{eden}
\epsilon(\sqrt{t^2-z^2}=\tau_0) = 
\frac{\epsilon_f}{\gamma^2}=\frac{\epsilon_f}{\cosh^2{\eta}}~,
\end{equation}  
where the space-time (pseudo)rapidity $\eta=\frac{1}{2}\ln{\left(\frac{t+z}{t-z}\right)}$ 
has been introduced. The first formula gives exactly the velocity field postulated in  
Bjorken's scaling hydrodynamics \cite{bjorken}, but here it follows from the conservation 
laws at the plasma boundary. The second equation shows that it is the global 
energy-density $T^{00}=\epsilon\gamma^2$ and not the rest-frame energy density 
$\epsilon$ which should be constant at the hypersurface $\tau=\tau_0$ where the 
plasma is created. As follows from eq. (\ref{eden}) the actual proper energy density
at $\tau=\tau_0$ decreases rather rapidly when going away from the central point $\eta=0$.
These predictions should be used as the initial conditions for 
further kinetic or hydrodynamical simulations of the plasma evolution.

The trajectories of the projectile and target slabs are not affected by the 
string decay until they intersect the hyperbola $\tau=\tau_0$. It is clear from the 
previous discussion that the region occupied by the plasma expands faster than the 
receding slabs. Thus the plasma eventually eats up the strings at time $t_a^{\ast}$ when the 
slab trajectories intersect with the hyperbola $\tau=\tau_0$, as shown in Fig.~2.
At this time the chromoelectric fields are fully neutralized and no new plasma 
is produced any more (see Fig.~1(c)). 
Substituting $z_a^{\ast}=z_a(t_a^{\ast})$
in eq. (\ref{cosh}) we obtain the final gamma-factors and rapidities of the slabs,
\begin{equation} \label{gamma}
\gamma_a^{\ast}=\cosh{y_a^{\ast}}==
\gamma_0\left[1-\frac{\tau_0}{\lambda_a}\left(v_0\sqrt{1+\frac{\tau_0^2}{4\lambda_a^2}}-
\frac{\tau_0}{2\lambda_a}\right)\right]~,
\end{equation}
where $\gamma_0=\cosh{y_0}$ and $v_0=\tanh{y_0}$.
We see that $\gamma_a^{\ast}$ is entirely determined by a single combination of 
parameters, $\tau_0/\lambda_a=(\epsilon_f\tau_0)/(\varepsilon_a\rho_0 l_a)$. 
Since $\varepsilon_a\simeq m_N$ is not expected to vary much in the course of the reaction 
and $l_a$ is given by the geometry, the combination $\epsilon_f \tau_0$ is the only essential 
parameter of the model which determines the rapidity loss by the nuclei. It is quite natural
that this is the same quantity which gives the initial transverse energy of the plasma at 
central pseudorapidity $\eta=0$. As follows from eq. (\ref{eden}), the energy density of the 
plasma at $\eta=0$ and $\tau=\tau_0$ is equal to $\epsilon_f$.
The total energy of the plasma in a slice $dz=\tau_0d\eta$ around $\eta=0$ is obtained by 
integrating $\epsilon_f({\bf b})$ over transverse area of the reaction zone. This gives
for central collisions
\begin{equation} \label{et}
\frac{dE_T}{d\eta}(\eta=0)=\pi R^2\langle\epsilon_f\rangle\tau_0~,
\end{equation}
where $\langle\epsilon_f\rangle$ is the energy density of the 
chromoelectric field averaged over the transverse area.

As our estimates show, it is most likely that $\lambda_a\ll\tau_0$.
In this case the partonic slabs are significantly moderated at the 
time $t_a^{\ast}$. Indeed, in this limit
$t_a^{\ast}\approx z_a^{\ast}\approx \lambda_a\gamma_0$, which is close to the turning 
point. From eq. (\ref{gamma}) the condition of complete stopping ($\gamma^{\ast}=1$) is
$\tau_0/\lambda_a=\sqrt{2(\gamma_0-1)}$. 
In collisions of asymmetric slabs the shorter slab may even 
reverse its motion in the direction of the longer slab, as illustrated in Fig.~2(b). 
Therefore, the proposed scenario provides an efficient mechanism for baryon stopping in 
ultrarelativistic heavy-ion collisions\footnote{ One should keep in mind, however, that 
even in the case of complete stopping the projectile and target slabs will be 
separated in space by a distance of about $(\lambda_p+\lambda_t)\gamma_0$.}. 
The baryon rapidity distribution is obtained by summing up contributions of all 
pairs of slabs and integrating over impact parameters. Our prediction is that in 
the first approximation this will be a superposition of two peaks 
centered at rapidities $y_p^{\ast}$ and $y_t^{\ast}$ given by eq. (\ref{gamma}). 
In a more refined approach one should also study what happens to the baryonic slabs after 
they are hit by the plasma. It is clear that the plasma wind will cause drift and diffusion 
of the baryon charge in rapidity space. As a result the projectile and target peaks will 
be shifted towards initial rapidities and smeared out. 

The process of plasma creation described above is similar to the breakdown of a 
dielectric in a strong electric field. The electromagnetic plasma produced in 
this case shows up as a spark or lightning. Since here we are dealing with the 
QCD plasma, this phenomenon can be referred to as ``QCD lightning''.     

A direct calculation of the energy density $\epsilon_f$ accumulated in the chromoelectric 
field at the early stage of a heavy-ion collision is problematic at present. Therefore, for 
our estimates we use a simple parameterization which is motivated by several model calculations.
\begin{equation} \label{eps}
\epsilon_f=\epsilon_0 \left(\frac{s}{s_0}\right)^{\alpha/2} 
\left(\frac{N_pN_t}{N_0^2}\right)^{\beta}~.
\end {equation}
Here the second factor is responsible for the incident energy dependence, $\sqrt{s}$ 
is the c.m. collision energy per nucleon pair. The exponent $\alpha\simeq 0.3$ 
follows from the low-$x$ behavior of the nuclear structure function or parton density 
\cite{kharzeev}.  The third factor accounts for the geometry of the overlap zone.
The exponent $\beta$ relates the number of strings produced to the number of binary
parton-parton collisions, which is proportional to $N_pN_t$. For convenience we have 
normalized $N_p$ and $N_t$ by the mean areal baryon density in the proton, 
$N_0\approx 0.5$ fm$^{-2}$. In the case of uncorrelated strings, as in elementary pp 
or p$\bar{p}$ collisions $\beta\approx1$, while in 
the case of strong string overlap or percolation $\beta\approx 0.5$ \cite{braun}. 
For the estimates below we use the value $\beta=0.5$, also used in ref.~\cite{csernai}, 
which seems appropriate for heavy-ion collisions at RHIC and LHC energies. It is 
assumed that this parameterization applies above a certain c.m. energy 
squared, $s\ge s_0$. The parameter $\epsilon_0$ can be determined at
$s=s_0$ and then used for higher energies.
With this choice we get 
\begin{equation} \label{l}
\frac{\tau_0}{\lambda_a} =\frac{\epsilon_f\tau_0}{\varepsilon_a\rho_0 l_a} \, \propto \,
\left(\frac{s}{s_0}\right)^{\alpha/2}\sqrt{\frac{l_{\tilde{a}}}{l_a}} ~,
\end{equation}
where $\tilde{a}=t$ for $a=p$ and vice versa. This formula shows that in an 
asymmetric slab-slab collision ($l_a<l_{\tilde{a}}$) the smaller slab will 
decelerate faster than the bigger one. This is a natural result since according 
to eq. (\ref{E}) their energy losses are equal. For equal slabs ($l_p=l_t=l$) the deceleration 
constant is independent of their lengths. This means that in a central collision 
the whole partonic sheet will decelerate uniformly. On the other hand, according to 
eq. (\ref{eps}), the energy density of the chromoelectric field decreases when going 
from the symmetry axis to the periphery, $\epsilon_f({\bf b})\propto l({\bf b})$.

We shall now discuss phenomenological implications of our model in light of recent 
RHIC data for central Au+Au collisions at $\sqrt{s}=130$ GeV ($y_0=4.94$). 
As found by the PHENIX collaboration \cite{phenix}, the transverse 
energy in the central pseudorapidity window is $dE_T/d\eta|_{\eta=0} \approx$ 
600 GeV. For the free streaming plasma ($p=0$) $\epsilon(\tau)\tau=\epsilon(\tau_0)\tau_0$, 
and $dE_T/d\eta|_{\eta=0}$ is independent of $\tau$. 
Then from eq. (\ref{et}) with $\pi R^2=148$ fm$^2$ 
we immediately obtain $\langle\epsilon_f\rangle\tau_0\approx 3.9$ GeV/fm$^2$.   
If the  pressure effects are fully included, $p = \epsilon/3$, the transverse energy drops 
as $\tau^{-1/3}$, and in order to obtain the observed $dE_T/d\eta$ one should increase 
the initial value  by a factor $(\tau_f/\tau_0)^{1/3}\simeq 2$, where 
$\tau_0$ and $\tau_f$ are the initial and final time of hydrodynamic expansion. 
In our estimates we take for $\langle\epsilon_f\rangle\tau_0$ the value 6 GeV/fm$^2$ which is 
in between these two extremes. This value is in qualitative agreement with results of 
microscopic simulations within the Quark-Gluon-String Model \cite{amelin}.

Now we can estimate the final rapidities of Au nuclei by considering the collision of 
representative slabs with lengths $l_p=l_t=\langle l\rangle=4R/3$ averaged over the transverse plane. 
Below we omit index $a$ for simplicity.  
The proper energy per baryon in partonic slabs can be approximated by
$m_T=\sqrt{m_N^2+\langle p_T\rangle^2}$, where 
$\langle p_T\rangle \approx$ 0.5 GeV/c as measured by the STAR Collaboration \cite{star}. 
To ensure energy conservation at $t=0$ we have reduced $\gamma_0$ by the factor $m_N/m_T$.
With these inputs we find from eq. (\ref{l}) that $\tau_0/\lambda=4.1$ or the characteristic deceleration 
length $\lambda=0.24\tau_0$ \footnote{It is interesting to note that with $\tau_0=1$ fm this
would correspond to the total deceleration time and distance $t^{\ast}\approx z^{\ast}\approx 14$ fm.}. 
From eq. (\ref{gamma}) we get  the final projectile and target  
rapidities, $y^{\ast}\approx \pm 1.9$. This corresponds to the final c.m. energy of 
about 3.5 GeV per baryon compared to the initial energy of 65 GeV per nucleon. 
The difference is transferred into the quark-gluon plasma.
This is a tremendous energy loss which can explain the  rather high degree of 
baryon stopping observed at RHIC. Unfortunately, the net-baryon rapidity distributions 
for this reaction are not available yet.  Preliminary BRAHMS data show that the net-proton rapidity 
distribution has a dip at $y=0$ and two peaks at $y = \pm (2\div 3)$ \cite{gaard}. 

Finally we use eq. (\ref{eps}) to make predictions for the full RHIC energy $\sqrt{s}=200 $ GeV.
Extrapolating from $\sqrt{s}=130$ GeV to $\sqrt{s}=200$ GeV, 
we get $\langle\epsilon_f\rangle\tau_0=6.8$ GeV/fm$^2$ and  $y^{\ast}=\pm 2.1$.
If instead we would use the values $\epsilon(\tau_0)=66.5$ GeV/fm$^3$ and $\tau_0=0.3$ fm 
given in ref.~\cite{krasnitz}, we would predict that
the baryonic slabs reach the turning point and reverse the direction of motion. 
Strict correlation between the transverse energy produced at central rapidities and the 
degree of baryon stopping is the most important prediction of our approach. 
The predicted strong deceleration of the baryon as well as the electric charge of ultrarelativistic
nuclei makes it very promising to search for photon bremsstrahlung at collider energies 
(see ref. \cite{kapusta} and references therein).

In conclusion, 
the collective dynamics of partonic slabs has been studied within a simple model 
incorporating strong chromoelectric fields generated early in the reaction. The 
model has two essential physical parameters: the energy density accumulated in 
the chromoelectric field, $\epsilon_f$, and the proper time required for its 
neutralization, $\tau_0$. Applying conservation of energy and momentum, we have 
established a direct correspondence between the rapidity shifts of the projectile 
and target nuclei and the transverse energy of the produced quark-gluon plasma. 
Using data from recent RHIC experiments we have estimated the energy 
density accumulated in the chromoelectric field and the resulting rapidity shift 
in the net baryon distribution.   

The authors thank P. J. Ellis and L. M. Satarov for useful discussions. I. N. M. 
acknowledges the kind hospitality of the Nuclear Theory Group, University of 
Minnesota. This research was supported by the Department of Energy under grant 
DE-FG02-87ER40328.

\vspace{1cm}

\begin{figure}[htp]
\centerline{\psfig{figure=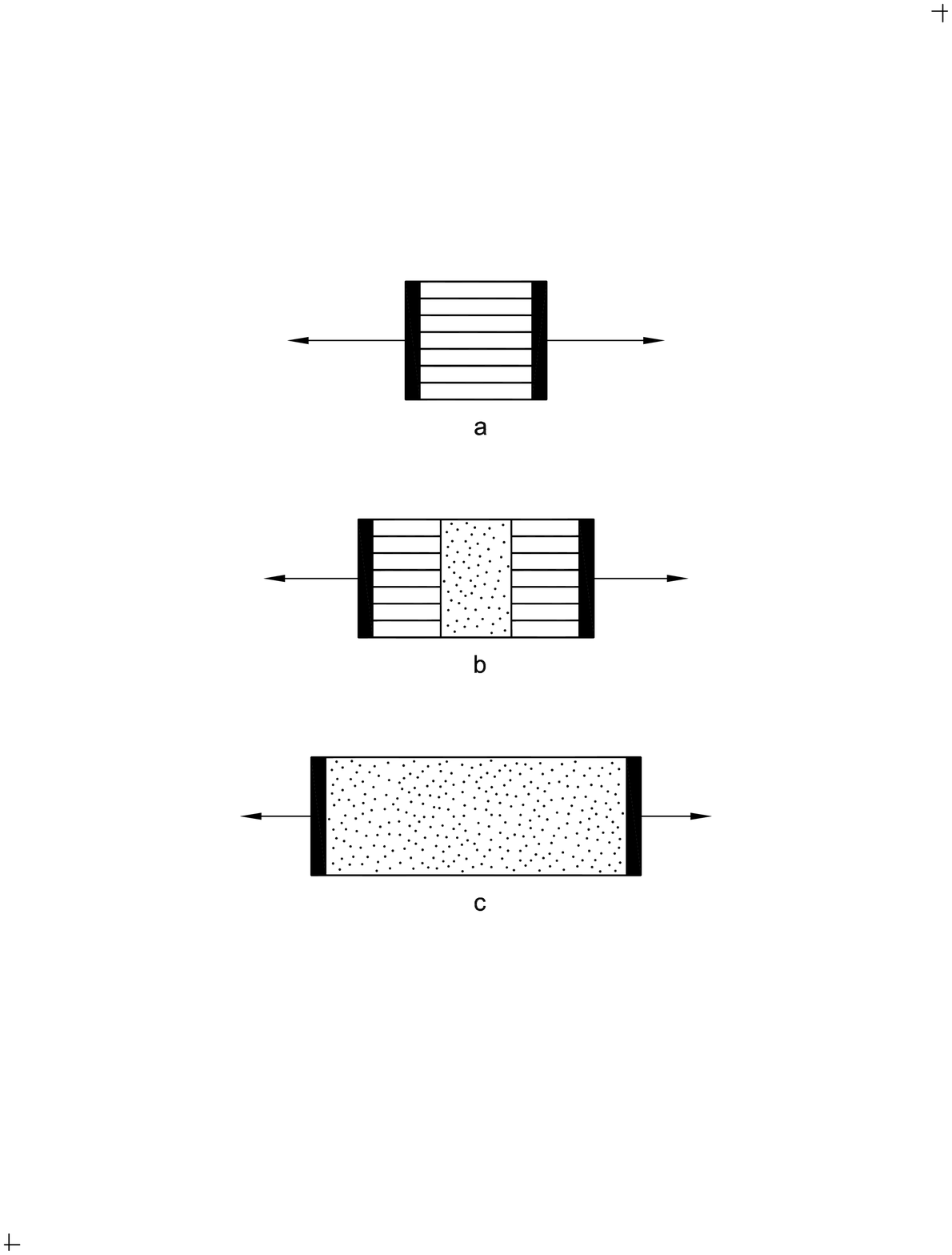,height=18.5cm}}
\caption{
Spatial view of a slab-slab collision at different times: 
(a) $0<t<\tau_0$, (b) $\tau_0<t<t^{\ast}$ and (c) $\tau\ge t^{\ast}$. The black 
boxes represent the receding projectile and target slabs. The horizontal 
straight lines indicate strings. The dotted areas show the regions occupied by 
the quark-gluon plasma.}
\end{figure}   

\begin{figure}
\centerline{\psfig{figure=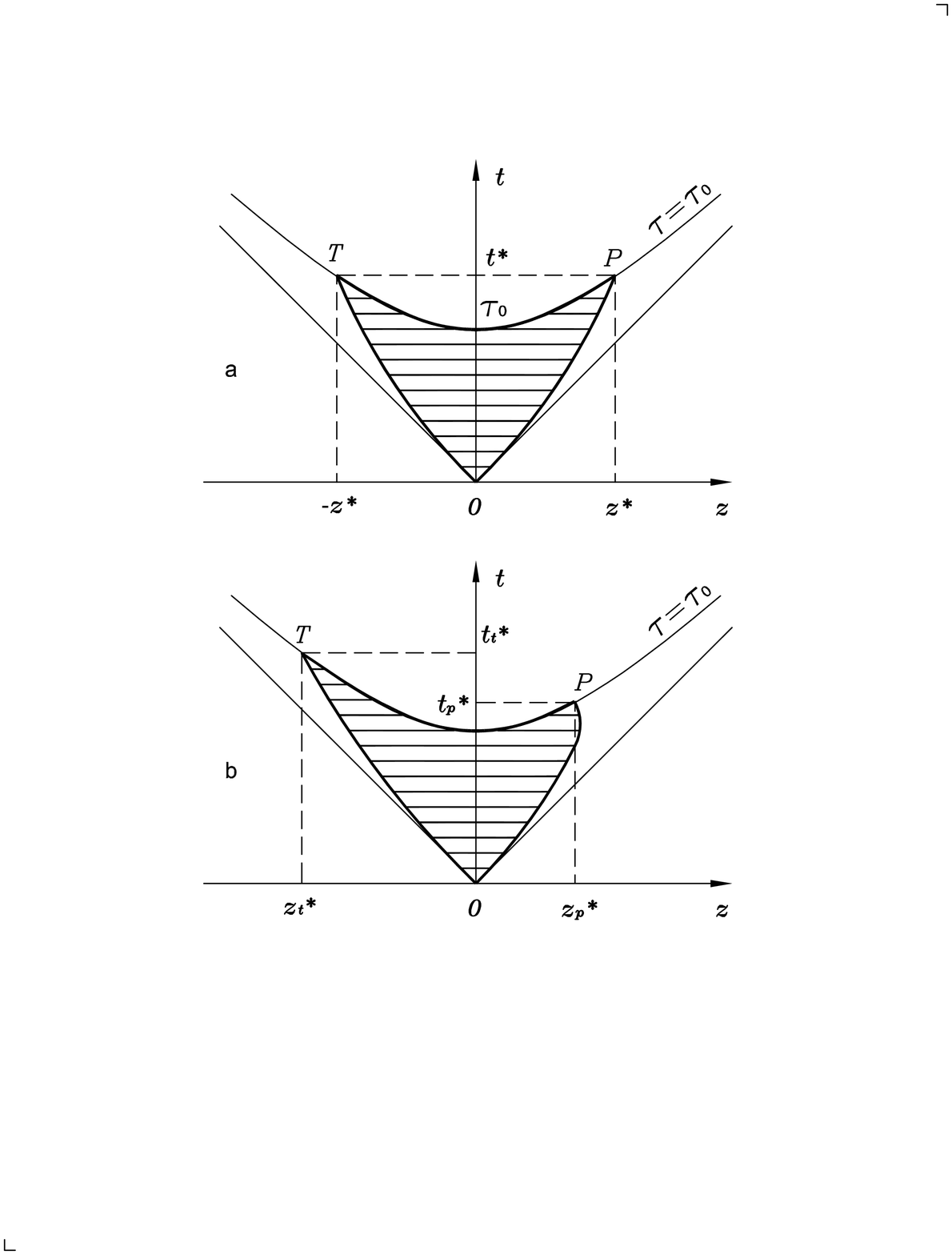,height=19cm}}
\caption{
Schematic space-time representation of a symmetric (a) and asymmetric (b) slab-slab collision 
in the $t-z$ plane. The projectile and target slab trajectories are shown by 
thick solid lines which start at the origin and terminate at points P and T, 
respectively. The quark-gluon plasma is produced at the portion of the hyperbola 
$\tau=\tau_0$ between these points (thick solid line).  Horizontal solid lines 
represent strings. For symmetric collisions $t_p^{\ast}=t_t^{\ast}=t^{\ast}$, 
$z_p^{\ast}=-z_t^{\ast}=z^{\ast}$.}
\end{figure}

\end{document}